# Coherent State Approach to Quantum Clocks

M. C. Ashworth [1]

*Department of Physics*
*University of California*
*Davis, California 95616 USA*

## Abstract

The "problem of time" has been a pressing issue in quantum gravity for some time. To help understand this problem, Rovelli proposed a model of a two harmonic oscillators system where one of the oscillators can be thought of as a "clock" for the other oscillator thus giving a natural time reference frame for the system. Recently, the author has constructed an explicit form for the coherent states on the reduced phase space of this system in terms of Klauder's projection operator approach. In this paper, by using coherent state representations and other tools from coherent state quantization, I investigate the construction of gauge invariant operators on this reduced phase space, and the ability to use a quantum oscillator as a "clock."

---

[1] *email: mikea@landau.ucdavis.edu*

# 1   Introduction

Coherent states have been a useful tool for studying quantum systems over the past several decades. Coherent states have several advantages over normal methods of quantization such as a natural relationship between quantum operators and classical observables, a well regularized path integral, and a natural geometrical structure. Recent work [1] - [4] has included constraints into this formulation. For constrained systems, coherent states offer still further advantages: the lack of gauge fixing, no Gribov ambiguities, and the ability to handle second class constraints without ambiguous determinants [2].

One example of a constrained systems is the time-reparameterization invariant double harmonic oscillator. The double harmonic oscillator is an important model in the study of quantum gravity [5], [6]. It is used in the investigation of the "problem of time." Classically, this system possesses a natural "clock" in terms of the motion of one of the oscillators. Even through the system is time reparameterization invariant, this "clock" can be used to define a natural time frame for the system. In the equations of motion (2.3), this is seen by replacing the time by the position of the second oscillator to write the position of the first oscillator without any direct time dependence,

$$q_1 = A \cos(\cos^{-1}(q_2/B) - \phi' + \phi). \tag{1.1}$$

So how do we write down an equivalent quantum operator for this idea the on reduced phase space? There has been much work towards making an meaningful quantum statement between the relations of these oscillators [7] - [9]. However, normal approaches to constructing these operators use stationary states where it is more difficult to understand a normal sense of time. It is natural to turn to coherent states to help with this problem. In fact, coherent states were first discovered in terms of a wave packet for the harmonic oscillator where the center moved with the classical frequency [10]. With this natural relation, it is possible to construct a "time reference frame" even at the quantum level.

In this paper, I will review the construction of the coherent state on the reduced phase space of the time-reparameterization invariant double harmonic oscillator [11]. In section 3, I will describe a projection scenario for operators. Because the momentum and position are placed on equal footing in a coherent state, it is possible to define a generic operator in terms of a diagonal representation or the "lower" symbol. Then using the projection operator on the states, this operator can be projected down to the reduced phase space. The unphysical degrees of freedom are then integrated over to give a well defined operator on the reduced phase space. In section 4, using a gauge invariant way of writing one function in terms of another considered by Marolf [8], I construct operators in terms a time reference frame from one of the oscillator. Then, I investigate the ability of this system to define this time reference frame at a quantum level.

## 2 Double harmonic oscillators

The double harmonic oscillator was first studied by Rovelli [5] as a model to help understand "the problem of time." Because of the advantages of coherent states, I would like to consider Rovelli's model in the frame work of coherent state quantization. In this section, I will review the construction of the coherent states on the reduced phase space of the double harmonic oscillator by means of the projection operator approach [11].

Let us start by considering the classical system. Let each of the harmonic oscillators have the same frequency, $\omega_1 = \omega_2 = \omega$. Then, the Hamiltonian for the time reparameterization invariant double harmonic oscillator is

$$H_T = \lambda \left( \frac{1}{2}(p_1{}^2 + \omega^2 q_1{}^2) + \frac{1}{2}(p_2{}^2 + \omega^2 q_2{}^2) - E \right), \tag{2.1}$$

and the action of this system is

$$S = \int p_1 dq_1 + p_2 dq_2 - \int H_T \, dt. \tag{2.2}$$

Because it is possible to absorb a reparameterization of the time coordinate into the Lagrange multiplier, $\lambda(t)$, this action is invariant under such reparameterizations. Moreover, the Lagrange multiplier is just a lapse function. The equations of motion are simplified in terms of the proper time, $\tau = \int_0^t \lambda(t)dt$,

$$\begin{aligned} q_1 &= A\cos(\omega\tau + \phi), & p_1 &= A\omega\sin(\omega\tau + \phi), \\ q_2 &= B\cos(\omega\tau + \phi'), & p_2 &= B\omega\sin(\omega\tau + \phi'). \end{aligned} \tag{2.3}$$

In addition to the equations of motion, the constraint equation must also be met,

$$\frac{1}{2}(p_1{}^2 + \omega^2 q_1{}^2) + \frac{1}{2}(p_2{}^2 + \omega^2 q_2{}^2) = E. \tag{2.4}$$

This constraint limits the amplitudes of the oscillators to

$$(A\omega)^2 + (B\omega)^2 = E. \tag{2.5}$$

To investigate the gauge degrees of freedom, let $\lambda \to \lambda + \varepsilon$, and then both angular coordinates are transformed as $\phi \to \phi + \varepsilon\omega t$ and $\phi' \to \phi' + \varepsilon\omega t$. The difference between the initial phases of the two harmonic oscillator, $\Delta\phi = \phi - \phi'$, is therefore gauge invariant. The two dimensional reduced phase space is then completely labeled by this phase difference and the ratio of the amplitudes of the oscillators.

On the full phase space, the coherent state is constructed in term of the standard set of Heisenberg operators, $\hat{P}_i, \hat{Q}^j$, where

$$[\hat{P}_i, \hat{Q}^j] = \delta_i^j, \qquad i, j = 1, 2. \tag{2.6}$$

Let us choose the fiducial vector to be the ground state of the quantum oscillators so that the system is physically centered.[1] Then the coherent state can be write as

$$|p, q\rangle = e^{-if(p,q)} e^{-\frac{i}{\hbar} p \cdot \hat{Q}} e^{\frac{i}{\hbar} q \cdot \hat{P}} |0\rangle. \tag{2.7}$$

This set of coherent state form an over complete set of vectors on the Hilbert space. In addition, for any choice of the fiducial vector, this representation admits a resolution of unity,

$$\mathbb{I} = \int |p, q\rangle \langle p, q| \prod_{j=1}^{N} \frac{dp_j dq^j}{2\pi}. \tag{2.8}$$

Now that we have the coherent state on the full phase space, let us next construct a set of coherent states on the reduced phase space.

There are two approaches we can use to construct a coherent state representation on the reduced phase space. The first is to construct a projection operator that will project the coherent states onto the physical states. The second approach is to find a set of appropriate operators that commute with the constraint operator and construct the coherent states from them. In this case, because the reduced phase space is spherical, the underlining symmetry is $SO(3)$, and the resulting operators are spin operators. For this system, the two approaches give an equivalent representation [11]. In general, however, finding such a set of operators may be difficult. Because the projection operator will be needed later, I will outline its construction here.

For convenience, it is easier to work in the complex coordinates,

$$\alpha = \sqrt{\frac{\omega}{2\hbar}} q_1 + \frac{i}{\sqrt{2\omega\hbar}} p_1, \tag{2.9}$$

$$\beta = \sqrt{\frac{\omega}{2\hbar}} q_2 + \frac{i}{\sqrt{2\omega\hbar}} p_2. \tag{2.10}$$

The position and moment operators are replaced with the standard raising and lower operators $(a, a^\dagger, b, b^\dagger)$. The constraint operator in terms of these operators can then be written as

$$\hat{\Phi} = a^\dagger a + b^\dagger b - E', \qquad \text{where} \qquad E' = E/\omega\hbar - 1. \tag{2.11}$$

The coherent state can be written in terms of the energy eigenstates,

---

[1] A coherent state is "physically centered" when $\langle p, q|\hat{P}_i|p, q\rangle = p_i$ and $\langle p, q|\hat{Q}^j|p, q\rangle = q^j$.

$$|\alpha, \beta\rangle = e^{-|\alpha|^2/2-|\beta|^2/2} \sum_{m,n}^{\infty} \frac{1}{\sqrt{n!}\sqrt{m!}} \alpha^m \beta^n |m, n\rangle. \qquad (2.12)$$

To project the coherent states on the physical states, let us use Klauder's projection operator construction [2]. Projecting the coherent states onto the physical states, gives

$$\begin{aligned} |\alpha, \beta\rangle_{phys} &= \mathbb{P}|\alpha, \beta\rangle \\ &= \int e^{i\lambda\hat{\Phi}} d\mu(\lambda) \left( e^{-|\alpha|^2/2-|\beta|^2/2} \sum_{m,n}^{\infty} \frac{1}{\sqrt{n!}\sqrt{m!}} \alpha^m \beta^n |m, n\rangle \right) \\ &= e^{-|\alpha|^2/2-|\beta|^2/2} \sum_{m,n}^{\infty} \frac{1}{\sqrt{n!}\sqrt{m!}} \alpha^m \beta^n \left( \int e^{i\lambda(n+m-E')} d\mu(\lambda) \right) |m, n\rangle \end{aligned} \qquad (2.13)$$

Then choosing a suitable measure (see [11] for more details), the physical vector is null unless $E'$ is arbitrarily close to an integer. Letting $E' = m' = m+n$, the physical vector is given by

$$|\alpha, \beta\rangle_{phys} = e^{-|\alpha|^2-|\beta|^2} \sum_{n=0}^{m'} \sqrt{\frac{1}{n!(m'-n)!}} \alpha^n \beta^{m'-n} |n, m'-n\rangle. \qquad (2.14)$$

Such a state is again a coherent state although not of the original Weyl-Heisenberg group but of a SO(3) group, as will be seen below. At this point, it is also easy to read off the orthonormal basis for the physical states from this expression. These are just

$$|\phi_n\rangle = |n, m'-n\rangle, \qquad n = 1, 2, 3, \ldots m', \qquad m' \text{ fixed.} \qquad (2.15)$$

This projected coherent state (2.14) still maintains the normalization from the full phase space and is not yet normalized on the reduced phase space. On the reduced phase space, the normalized coherent state is

$$|\alpha, \beta\rangle'_{phys} = \left(|\alpha|^2 + |\beta|^2\right)^{-\frac{m'}{2}} \sum_{n=0}^{m'} \sqrt{\frac{m'!}{n!(m'-n)!}} \alpha^n \beta^{m'-n} |n, m'-n\rangle. \qquad (2.16)$$

The gauge transformation of this system transforms the complex coordinates as $\alpha \to \alpha e^{i\theta}$ and $\beta \to \beta e^{i\theta}$. In the coherent state, these gauge transformation appears as an overall phase in front of the physical vector,

$$|\alpha, \beta\rangle'_{phys} \to e^{im'\theta} |\alpha, \beta\rangle'_{phys}. \qquad (2.17)$$

Let us define the complex coordinate, $\xi = \alpha/\beta$, which is independent of the gauge transformation. The real part of this coordinate is the ratio of the amplitudes of the harmonic oscillators, and the complex phase gives the phase difference between the oscillators. Thus this coordinate completely labels the reduced phase space. Factoring out the gauge transformations, the coherent state is

$$
\begin{aligned}
|\alpha, \beta\rangle_{phys} &= \left(\frac{\beta}{|\beta|}\right)^{m'} \left(1 + \left|\frac{\alpha}{\beta}\right|^2\right)^{-\frac{m'}{2}} \sum_{n=0}^{m'} \sqrt{\frac{m'!}{n!(m'-n)!}} \left(\frac{\alpha}{\beta}\right)^n |n, m'-n\rangle \\
&= e^{im'\theta} \left(1 + |\xi|^2\right)^{-\frac{m'}{2}} \sum_{n=0}^{m'} \sqrt{\frac{m'!}{n!(m'-n)!}} \xi^n |n, m'-n\rangle \\
&= e^{im'\theta}|\xi\rangle.
\end{aligned}
\qquad (2.18)
$$

The physical coherent state then maps onto the $SO(3)$ coherent state [12], where the energy is mapped onto the total angular momentum, $2j = m'$,

$$
|\xi\rangle = (1 + |\xi|^2)^{-j} \sum_{m=-j}^{j} \sqrt{\frac{2j!}{(j+m)!(j-m)!}} \xi^{j+m} |j, m\rangle. \qquad (2.19)
$$

Now that we have the coherent states on the reduced phase space, the next step is to find the resolution of unity. For this case, we have a standard $SO(3)$ coherent state representation for which the resolution of unity is already known [13]. However, it is also possible to construct the resolution of unity by means of the projection operator. From the resolution of unity on the full phase space, we can project the unity operator to find the unity operator on the reduced phase space,

$$
\mathbb{I}' = \mathbb{P}\mathbb{I}\mathbb{P} = \int \mathbb{P}|\alpha, \beta\rangle\langle\alpha, \beta|\mathbb{P} \left(\frac{d\alpha d\bar{\alpha}}{\pi}\right) \left(\frac{d\beta d\bar{\beta}}{\pi}\right). \qquad (2.20)
$$

Then substituting the above definition of the physical vector,

$$
|\xi\rangle = |\alpha, \beta\rangle'_{phys} = \frac{\mathbb{P}|\alpha, \beta\rangle}{||\mathbb{P}|\alpha, \beta\rangle||}, \qquad (2.21)
$$

for the project states, the resolution unity looks like

$$
\mathbb{I}' = \int |\xi\rangle\langle\xi| \left|\langle\alpha, \beta|\mathbb{P}|\alpha, \beta\rangle\right| \left(\frac{d\alpha d\bar{\alpha}}{\pi}\right) \left(\frac{d\beta d\bar{\beta}}{\pi}\right). \qquad (2.22)
$$

Then changing to the coordinates,

$$
r = |\alpha|^2 + |\beta|^2,
$$

$$e^{i\theta} = \beta/|\beta|, \quad \text{where} \quad (0 < \theta < 2\pi),$$
$$\xi = \alpha/\beta, \tag{2.23}$$

the resolution of unity becomes

$$I\!\!I' = \int |\xi\rangle\langle\xi| \frac{d\xi d\bar{\xi}}{\pi(1+|\xi|^2)^2} \left(\frac{e^{-r}r^m}{m!}\right) \frac{rdrd\theta}{2\pi}. \tag{2.24}$$

Integrating over the constraint coordinate $r$ and the gauge orbit $\theta$ removes their dependence. Note that the integrand is strongly peaked at the constraint surface and that it is independent of the gauge orbit. The result, after integrating, is the standard form of the $SO(3)$ coherent state resolution of unity,

$$I\!\!I' = \frac{(2j+1)}{\pi} \int |\xi\rangle\langle\xi| \frac{d\xi d\bar{\xi}}{(1+|\xi|^2)^2}. \tag{2.25}$$

Now that we have the coherent state and the resolution of unity, we can consider the dynamics of this system. Because the reduced Hamiltonian is zero, there is no "time" evolution of this system. The resulting "propagator" is simply the overlap function of the $SO(3)$ coherent state,

$$\begin{aligned}
\langle \xi'|\xi\rangle &\equiv \frac{\langle \alpha',\beta'|I\!\!P|\alpha,\beta\rangle}{|\langle \alpha',\beta'|I\!\!P|\alpha',\beta'\rangle|\, |\langle \alpha,\beta|I\!\!P|\alpha,\beta\rangle|} \\
&= (1+|\xi'|^2)^{-j}(1+|\xi|^2)^{-j}(1+\bar{\xi}'\xi)^{2j}.
\end{aligned} \tag{2.26}$$

So how can we get a sense of "time" out of such a system? To answer this question, let us turn to the set of operators that are well defined on the reduced phase space.

## 3  Projection of operators

Classically, it is clear how to project a function onto the reduced phase space. Using the natural relation between operators and classical functions from a coherent state picture, it is possible to construct a projected operator. This relationship between operator and their corresponding classical function is realized in terms of symbols. In coherent state quantization, there are two natural definitions of symbol of an operator [13]. The "upper" symbol, $O(p,q)$, is just the expectation value of the operator,

$$\begin{aligned}
O(p,q) &= \langle p,q|\tilde{O}(\hat{P},\hat{Q})|p,q\rangle \\
&= \langle \eta|\tilde{O}(\hat{P}+pI\!\!I,\hat{Q}+qI\!\!I)|\eta\rangle \\
&= \tilde{O}(p,q) + \mathcal{O}(\hbar),
\end{aligned} \tag{3.1}$$

where $|\eta\rangle$ is the physically centered fiducial vector. The "lower" symbol, $o(p,q)$, is defined implicitly in terms of the diagonal representation of the operator,

$$\tilde{O}(\hat{P},\hat{Q}) = \int o(p,q)\,|p,q\rangle\langle p,q|d\mu(p,q). \tag{3.2}$$

Certainly not all operators will have a well defined diagonal representation (see [13]). However, for a reasonable large set of operators, such a representation is possible.

After projecting to the reduced phase space, not all operators are well defined on the physical states. For example, the raising operator for one of the harmonic oscillator acting on a physical state (2.15),

$$a|n,m-n\rangle = |n,m-n+1\rangle, \tag{3.3}$$

does not give back a physical state (2.15). In addition, an observable operator, ie an Hermitian operator on the full phase space, may have gauge dependence when acting on a physical vector. Therefore, the upper symbol is only well defined on the reduced phase for operators that take physical states to physical states and are inherently gauge invariant. In which case, the symbol may be written as

$$O(p,q)\Big|_{phys} = \frac{\langle p,q|\mathbb{P}\,\tilde{O}(\hat{P},\hat{Q})\,\mathbb{P}|p,q\rangle}{|\langle p,q|\mathbb{P}|p,q\rangle|}. \tag{3.4}$$

So at this point, it becomes a question of finding such well behaved operators. In terms of the double harmonic oscillator, an example of such operators are the spin operators $S_i$ (see [5], [11] for the construction of these operators). These operator commute with the constraint operator (2.11) and are well defined on the physical states. The following are the well defined upper symbols or expectation values for these operators:

$$\begin{aligned} s_1 = \langle\xi|S_1|\xi\rangle &= \frac{2j\mathrm{Re}[\xi]}{(1+|\xi|^2)}, \\ s_2 = \langle\xi|S_2|\xi\rangle &= \frac{2j\mathrm{Im}[\xi]}{(1+|\xi|^2)}, \\ s_3 = \langle\xi|S_3|\xi\rangle &= -j\frac{(1-|\xi|^2)}{(1+|\xi|^2)}. \end{aligned} \tag{3.5}$$

On the other hand, we may take a different approach. For the "lower" symbol, the situation is reversed, the operator is defined in terms of the symbol (3.2). From this definition, it is possible to project an operator onto a well defined operator on the physical states by projecting the states in the diagonal representation,

$$\tilde{O}(\hat{P},\hat{Q})\Big|_{phys} = \int o(p,q)\,\mathbb{P}|p,q\rangle\langle p,q|\mathbb{P}\,d\mu(p,q). \tag{3.6}$$

This operator now appears in the same form as the identity operator (2.20). Like the identity operator, it is possible to integrate out the unphysical degrees of freedom leaving a new diagonal representation in terms of the reduced phase space coherent states. To see this in more detail, let us work through the example of the double harmonic oscillator.

Repeating the process from the projection of the resolution of unity (2.20), for a general operator,

$$\begin{aligned}\tilde{O}(\hat{P},\hat{Q})\Big|_{phys} &= \int o(\alpha,\beta)\,I\!\!P|\alpha,\beta\rangle\langle\alpha,\beta|I\!\!P\left(\frac{d\alpha d\bar{\alpha}}{\pi}\right)\left(\frac{d\beta d\bar{\beta}}{\pi}\right)\\ &= \int o(\alpha,\beta)\,|\xi\rangle\langle\xi|\big|\langle\alpha,\beta|I\!\!P|\alpha,\beta\rangle\big|\left(\frac{d\alpha d\bar{\alpha}}{\pi}\right)\left(\frac{d\beta d\bar{\beta}}{\pi}\right)\\ &= \frac{(m+1)}{\pi}\int o'(\xi)\,|\xi\rangle\langle\xi|\frac{d\xi d\bar{\xi}}{(1+|\xi|^2)^2},\end{aligned} \qquad (3.7)$$

where $o'(\xi)$ is the "projected symbol,"

$$o'(\xi) = \frac{1}{(m+1)}\int o(\alpha,\beta)\left(\frac{e^{-r}r^m}{m!}\right)\frac{rdrd\theta}{2\pi}. \qquad (3.8)$$

This operator is well defined on the physical states. As our next step, let us take a closer look at this projected symbol.

In the direction of the constraint, $r$, note that the integrand function (expanding about the maximum),

$$\frac{e^{-r}r^{m+1}}{m!} \sim \frac{(m+1)}{\sqrt{2\pi(m+1)}}\exp\left\{\frac{(r-(m+1))^2}{2(m+1)}\right\}, \qquad (3.9)$$

is a strongly peaked function at $r = m+1$, the constraint surface (2.5). In fact, in the classical limit, $\hbar \to 0$, this function becomes a delta function. So to first order in $\hbar$, the projection of the lower symbol is restricted to the constraint surface,

$$o'(\xi) = \int\left(\frac{d\theta}{2\pi}o(\alpha,\beta)\Big|_{r=m+1}\right) + \mathcal{O}(\hbar). \qquad (3.10)$$

Next, we note that $\theta$ is the parameter of the gauge orbit. If the lower symbol is not dependent on this parameter, ie is gauge independent, then integrating over $\theta$ is just unity and

$$o'(\xi) \approx o(\alpha,\beta)\Big|_{r=m+1}. \qquad (3.11)$$

If there is dependence on the gauge orbit, then the resulting lower symbol is the average over the gauge group, which is a natural definition of a projection operator [9].

Looking at an important example of a projected operator, let us consider the position operator, $\hat{Q}_2$. Its symbol is dependent on the gauge orbit, $\theta$,

$$\begin{aligned} \hat{Q}_2 &= \int \left(\frac{\beta + \bar{\beta}}{\sqrt{2}}\right) |\alpha, \beta\rangle\langle\alpha, \beta| \left(\frac{d\alpha d\bar{\alpha}}{\pi}\right) \left(\frac{d\beta d\bar{\beta}}{\pi}\right) \\ &= \int \frac{|\beta|}{\sqrt{2}} \cos\theta \, |\alpha, \beta\rangle\langle\alpha, \beta| \left(\frac{d\alpha d\bar{\alpha}}{\pi}\right) \left(\frac{d\beta d\bar{\beta}}{\pi}\right). \end{aligned} \quad (3.12)$$

Then the reduced lower symbol is of the form,

$$q_2'(\xi) \sim \int \cos\theta \frac{d\theta}{2\pi} = 0. \quad (3.13)$$

The fact that this symbol is zero should not come as a surprise. The gauge transformation in this system are really "time" translations. The resulting averaging over the gauge orbits, results in a time average over one period of the oscillator. So the average position is zero.

## 4 The coherent state oscillators as a ideal clock

As was stated earlier, the position of the first oscillator can be written in terms of the position of the second oscillator,

$$q_1 = A\cos(\cos^{-1}(q_2/B) - \phi' + \phi). \quad (4.1)$$

However, this function is still dependent on the gauge transformation. Placing this directly into the projection scheme above, still leads to a null operator (3.13). In order to use the second oscillator as a clock for a time reference frame, another function must be considered. Marolf [8] noted that, this classical statement can be replaced by an equivalent gauge invariant statement,

$$o\Big|_{q=\tau} = \int dt \frac{dq}{dt} \delta(q(t) - \tau) o(t). \quad (4.2)$$

This function is now reparameterization invariant and can be used to construct a meaningful operator on the reduced phase space.

Because $\theta$ in our choice of coordinates is the phase of the second harmonic oscillator (2.23), it may be considered to be the time of this system. Thus it is possible to insert this form of the delta function above into our definition of the project symbol (3.8),

$$o'(\xi)\Big|_{q=\tau} = \int o(\xi,r,\theta)\left(\frac{e^{-r}r^{m+1}}{(m+1)!}\right)\frac{dq}{d\theta}\delta(q(\theta)-\tau)\frac{drd\theta}{2\pi}. \tag{4.3}$$

The delta function gives a particular gauge slice or time frame for the system. So returning to our original idea of measuring the one oscillator in terms of the other (1.1), we can define a time by the equation of motion of the second oscillator,

$$q_2(\theta) = B\cos(\omega\tau + \phi'). \tag{4.4}$$

Placing this into (4.3),

$$\begin{aligned}q'_1(\xi)\Big|_{q_2=\tau} &= \int q_1(\xi,r,\theta)\frac{e^{-r}r^{m+1}}{(m+1)!}\frac{dq_2}{d\theta}\delta\big(q_2(\theta) - B\cos(\omega\tau+\phi')\big)\frac{drd\theta}{2\pi}\\ &= \int\left(\xi e^{i\theta} + \bar{\xi}e^{-i\theta}\right)\sqrt{\frac{r}{(1+|\xi|^2)}}\frac{e^{-r}r^{m+1}}{(m+1)!}\delta(\theta-(\omega\tau+\phi'))\frac{drd\theta}{2\pi}\\ &= \frac{\Gamma(m+5/2)}{(m+1)!}\frac{\xi e^{i(\omega\tau+\phi')} + \bar{\xi}e^{-i(\omega\tau+\phi')}}{(1+|\xi|^2)^{\frac{1}{2}}}. \end{aligned} \tag{4.5}$$

In the limit where the energy becomes large ($m \to \infty$), this symbol becomes

$$q'_1 = A\cos(\omega\tau + \phi' + \delta\phi), \tag{4.6}$$

where $A$ is the amplitude of the first oscillator and $\delta\phi$ is the phase difference between the oscillators. This is the classical equation of motion (2.3). For this reduced symbol, the corresponding operator is defined,

$$Q'_1 = \int q'_1(\xi)\,|\xi\rangle\langle\xi|d\mu(\xi). \tag{4.7}$$

Because in the classical limit the two symbols are equivalent, the expectation value of this operator also leads to the classical equation of motion.

Using the projection of the symbols, we are able to construct meaningful operators on the reduced phase space. Then by using a gauge invariant function (4.2), we can construct an operator in terms of the position of one of the oscillators. This operator in the classical limit gives the equation of motion. Therefore we can use such an operator to establish a time reference frame.

The next question is how well this time frame works on a quantum level? To answer this question, we will return to the propagator on the reduced phase space (2.26),

$$\begin{aligned}\langle\xi'|\xi\rangle &\equiv \langle\alpha',\beta'|\mathbb{P}|\alpha,\beta\rangle\\ &= (1+|\xi'|^2)^{-j}(1+|\xi|^2)^{-j}(1+\bar{\xi}'\xi)^{2j}. \end{aligned} \tag{4.8}$$

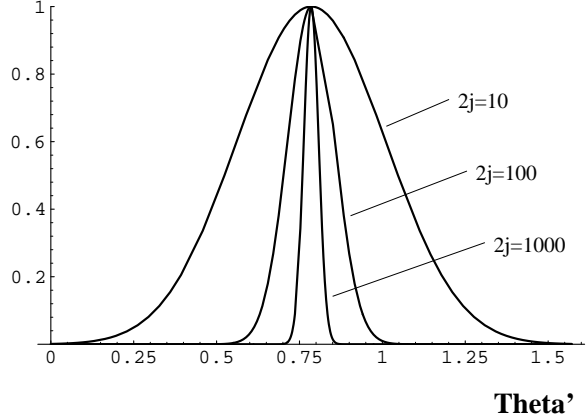

Figure 1: The correlation function with respect to $\Theta'$ with $\Theta = \pi/2$.

The magnitude of $\xi$ gives the ratio of the amplitudes of the two harmonic oscillators. Let this ratio be given in terms of a angle,

$$\tan\Theta = |\xi| = \frac{|A|}{|B|}. \tag{4.9}$$

The coherent state is not an eigenstate of the corresponding operator, so there is some overlap between states with different amplitude ratios. However, the correlation between different coherent state is strongly peaked when this angle is preserved. In fact, it is a Gaussian about the initial angle that becomes shaper at higher energies (see figure 1). The width of this Gaussian is

$$\sigma^2 = \frac{1}{2j} \tag{4.10}$$

So this function becomes quickly peaked in the high energy limit.

Let us turn to the other source of information in the correlation function, the imaginary phase difference between the initial and final states, $\text{Im}[\xi\bar{\xi}'] = \cos\delta\phi$. In the classical limit, $2j$ becoming large, and the transition amplitude is peaked at $\delta\phi = 0$. This means that the two oscillators stay in the same relative phase with each other, which is of course true classically. The correlation function between the phases is approximately Gaussian (see figure 2). The width of this Gaussian is

$$\sigma^2 = \frac{2j}{E_1 E_2}, \tag{4.11}$$

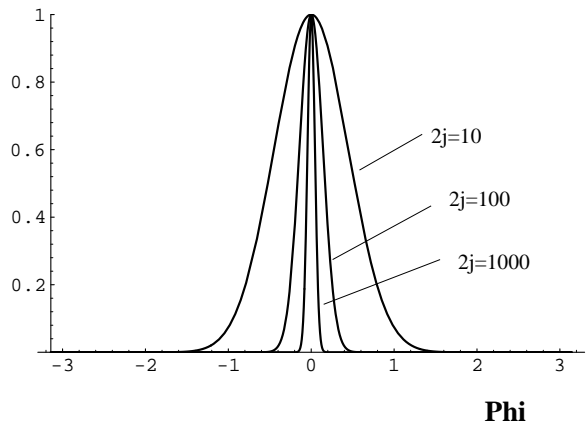

Figure 2: The correlation function of $\delta\phi$ - the "the clock error."

where $E_1$ and $E_2$ are the energy for each oscillator. If the total energy is roughly shared by each of the oscillator, then width also becomes sharply peaked. So at a classical level, there is a meaningful sense of time as measured by the clock. However at low energies, although there is a well defined time, the motion of the two oscillators are not synchronized. In terms of this time frame, it is possible that between two points in time the first oscillator may seem to be moving backwards.

Using coherent states, it is possible to use the natural relation between the "lower" symbol and the quantum operator to project an operator onto the reduced phase space. Again using the natural relation between operators and classical functions, it is possible to construct an operator based on the position of one of the operators. This construction gives a natural "time reference frame" that corresponds to a classical time. However, in a quantum energy regime, the "quantum clock" seems to be a poor time keeper.

# 5 Acknowledgments

I would like to thank Steve Carlip for all of his support and time, and John Klauder for his comments. This work was supported by National Science Foundation grant PHY-93-57203 and Department of Energy grant DE-FG03-91ER40674.